# Exploiting plume structure to decode gas source distance using metal-oxide gas sensors

Michael Schmuker[1,2,3] *, Viktor Bahr[2], Ramón Huerta[3]

1: University of Sussex, School of Engineering and Informatics, Falmer, Brighton BN1 9QJ, UK.

2: Freie Universität Berlin, Dept. of Biology, Chemistry, Pharmacy, Königin-Luise-Str. 1-3, 14195 Berlin, Germany

3: University of California San Diego, BioCircuits Institute, La Jolla, CA 92093-0328, USA.

* corresponding author

**Abbreviations:**

| | |
|---|---|
| MOX sensor | — Metal-Oxide sensor |
| PID | — Photo-Ionisation detector |
| $RMSE_{(CV)}$ | — Root-mean-square error (in cross-validation) |
| IBI | — Inter-bout intervals |

# Abstract

Estimating the distance of a gas source is important in many applications of chemical sensing, like e.g. environmental monitoring, or chemically-guided robot navigation. If an estimation of the gas concentration at the source is available, source proximity can be estimated from the time-averaged gas concentration at the sensing site. However, in turbulent environments, where fast concentration fluctuations dominate, comparably long measurements are required to obtain a reliable estimate. A lesser known feature that can be exploited for distance estimation in a turbulent environment lies in the relationship between source proximity and the temporal variance of the local gas concentration – the farther the source, the more intermittent are gas encounters. However, exploiting this feature requires measurement of changes in gas concentration on a comparably fast time scale, that have up to now only been achieved using photo-ionisation detectors. Here, we demonstrate that by appropriate signal processing, off-the-shelf metal-oxide sensors are capable of extracting rapidly fluctuating features of gas plumes that strongly correlate with source distance. We show that with a straightforward analysis method it is possible to decode events of large, consistent changes in the measured signal, so-called 'bouts'. The frequency of these bouts predicts the distance of a gas source in wind-tunnel experiments with good accuracy. In addition, we found that the variance of bout counts indicates cross-wind offset to the centreline of the gas plume. Our results offer an alternative approach to estimating gas source proximity that is largely independent of gas concentration, using off-the-shelf metal-oxide sensors. The analysis method we employ demands very few computational resources and is suitable for low-power microcontrollers.

# 1 Introduction

Estimating the distance of a gas source is important in many scenarios. For example, when monitoring environmental concentrations of certain gases with a stationary sensor, an estimate of the distance to the source of gas emission will help in localising the emission site. Likewise, the success of a robotic agent trying to localise the source of a hazardous gas leak will depend on the speed and accuracy of its estimates of how far upwind the source is. In the biological realm, gas-based navigation plays a crucial role in insects and mammals looking for food and mating partners, or trying to avoid predators.

A prime clue to the distance of a gas source is the concentration of the gas. Downwind from the source the gas concentration will decrease through diffusion and the gas plume will be mixed with air by turbulent (advective) processes [1]. If the gas concentration at the source and the wind speed are known, the distance to the source can, in theory, be estimated from the amount of dilution that has taken place while the gas filament has travelled to the site of detection. But this estimation method contains many sources of error, like e.g. an unknown concentration at the source, as in field applications like the localisation of gas leaks or fires. Moreover, in a turbulent environment the measured concentration of a gas released at a remote upwind site highly intermittent [2]. For reliable estimates of gas concentration, measurements have to be averaged over a time interval, prolonging the procedure. Experimental evidence even suggests that in a turbulent, uncontrolled environment, the average concentration is not a very good estimator of source distance, and that the variance of concentration yields better distance estimates [3].

Interestingly, the intermittent nature of a gas plume itself contains information about source distance that is independent of concentration. It has been shown that the rate of concentration fluctuation in a turbulent environment correlates well with the distance to the gas source in the open field [2,4] and in wind-tunnel experiments [5,6]. Generally, fast fluctuations dominate the signal close to the source, while slower components become more prominent further away.

These experimental observations were made using photo-ionisation detectors (PIDs), which can resolve changes in gas concentration on the order of 100 Hz or more [4,6]. However, their high cost and complexity prohibit their use in low-cost, portable solutions for gas sensing. In addition, PIDs typically don't differentiate well between gas species. Hence, PIDs alone will not be sufficient if one is interested not only in resolving gas concentration at a reasonable temporal resolution, but also in identifying gases. If gas identification is important, PIDs have to be supplemented with another technology that better supports gas discrimination.

Metal-Oxide (MOX) gas sensors provide reasonable accuracy in identifying single gases [7] and mixtures [8]. They can be obtained at low cost, are easily integrated into electronic circuits, and they draw comparably little power. However, MOX-type sensors suffer from limited resolution in the temporal domain. While the initial response has been found to be quite fast (with a time constant of 1 to 10 $s$ depending on the sensor parameters, target gas and other conditions), the recovery time constant is comparably slow (around 100 $s$) [9]. Due to their fast response onset and slow recovery, MOX sensors effectively act as "leaky integrators" of gas concentration. Therefore, in a turbulent environment where gas concentration is quickly fluctuating, the temporal resolution of MOX sensors is likely limited by the component of the signal that is caused by sensor recovery, which overshadows the fast transient initial response.

This limitation can be mitigated by signal processing methods [10,11]. The sensor response can be represented by a simple exponential model, that allows to predict the steady-state response from the transient phase of the MOX-sensor response [12,13]. This approach essentially consists of applying a bandpass filter, as the signal is first differentiated (high-pass) and then convolved with an exponential kernel ("leaky integration", low-pass). This method has also been used to improve the detection of change-points in gas concentrations in a open-sampling environment [14]. These studies indicate that in spite of their reputation of responding slowly to changes in gas concentration, MOX-sensors do indeed provide information that is encoded on comparably fast time scales, which can be extracted by appropriate signal processing.

Our aim in this study was to analyse whether off-the-shelf MOX-sensors can reveal fluctuations in gas concentration that can be exploited to predict the distance of the sensor from the gas source. We based our analysis on a large wind tunnel dataset provided by Vergara and colleagues [7]. We first show, using a band-pass approach, that the sensor signal contains information about fast fluctuations that is sufficient to predict source distance. We then go on to increase the precision of distance estimation by more elaborate signal processing. We analyse how this prediction method depends on the gas species, wind speed, and sensor parameters. Finally, we show that the fast fluctuations extracted by a straightforward cascading differentiation/integration approach are equivalent to de-filtering the signal by de-convolution with the sensor's impulse response.

## 2   Materials and Methods

### 2.1   Data

This study is based on a public dataset recorded by Vergara and coworkers [7], that contains recordings from MOX-type sensors in a wind-tunnel. Sensor boards were placed at one of six discrete distances from a gas outlet ($0.25\ m$, $0.5\ m$, $0.98\ m$, $1.18\ m$, $1.40\ m$, $1.45\ m$). Figure 1 shows a schematic of the wind tunnel and the positions of the sensor boards.

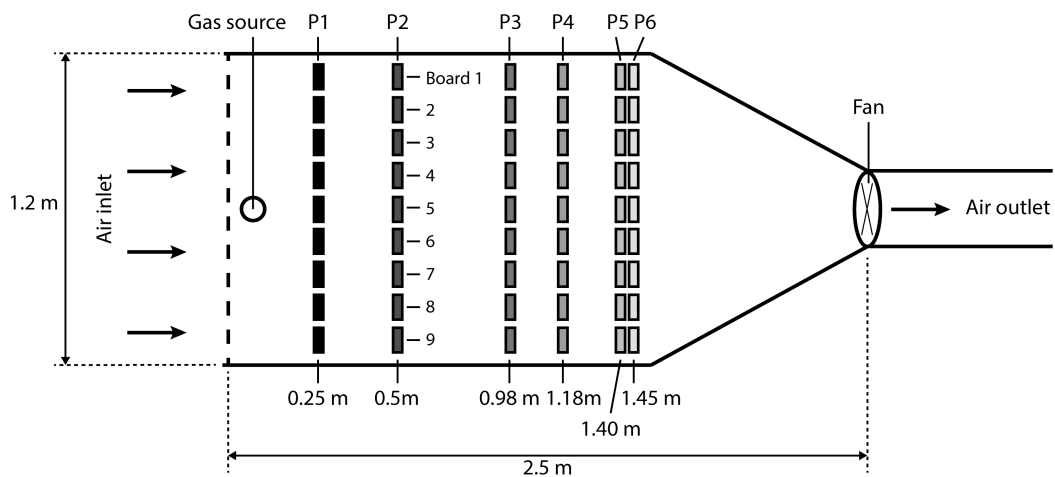

**Fig. 1:** Schematic of the wind tunnel set-up (adapted and modified from [7]). The position of the sensor boards relative to the gas source is coded in the grey value of the board symbols, as well as in the signal traces in the following figures.

Each individual board was equipped with a total of 8 sensors from Figaro Inc., models TGS 2600 (2x), TGS 2602, TGS 2610, TGS 2611, TGS 2612, and TGS 2620 (2x). Table 1 lists the mapping between sensor models and columns in the data set.

**Table 1:** Mapping data columns to sensor models.

| Col. no. | Sensor model | Col. no. | Sensor model |
|---:|---|---:|---|
| 1 | TGS 2611 | 5 | TGS 2600 (a) |
| 2 | TGS 2612 | 6 | TGS 2600 (b) |
| 3 | TGS 2610 | 7 | TGS 2620 (a) |
| 4 | TGS 2602 | 8 | TGS 2620 (b) |

Recordings were obtained at one of three wind speeds ($0.10\ m/s$, $0.21\ m/s$ and $0.34\ m/s$), and at one of six voltages controlling the sensor's heating element (between 4.0 and 6.0 V). Unless stated otherwise, we used the signals acquired along the centreline of the wind-tunnel, i.e. from board no. 5.

Sensor 1 was excluded from the analysis since it often contained artefacts (at least on Board 5), which could not be mitigated by post-processing (see supplemental Fig. S1 and accompanying text).

A total of 10 gases have been sampled: Acetaldehyde (2500 $ppm$), Acetone (500 $ppm$), Ammonia (10000 $ppm$), Benzene (200 $ppm$), Butanol (100 $ppm$), Carbon Monoxide (CO, 1000 $ppm$ and 4000 $ppm$), Ethylene (500 $ppm$), Methane (1000 $ppm$), Methanol (200 $ppm$), Toluene (200 $ppm$).

The full setup, including a detailed description of signal acquisition and an initial analysis are described in the original publication [7]. The full dataset (approx. 8 GB) is available at https://archive.ics.uci.edu/ml/datasets/Gas+sensor+arrays+in+open+sampling+settings .

## 2.2  Bout detection algorithm

We used a cascaded filtering approach to enhance fast transients in the signal [12,13], and subsequently detected "bouts" in the signal, i.e. portions where the amplitude of the filtered signal was consistently rising. The signal $S$ was first smoothed (i.e., low-pass filtered) by convolution with a Gaussian kernel to remove high-frequency noise (eq. 1),

$$S_{smooth} = S * G^\sigma \tag{1}$$

We used a kernel $G^\sigma$ with $\sigma_{smooth} = 0.3$ s for all signals, except in section 3.7, where we explored the impact of changing the kernel parameters.

We then formed the derivative of the signal (difference of sequential data points, as implemented in the function `numpy.diff()` in python), as in equation 2:

$$x_t = s_t - s_{t-1}, \tag{2}$$

with $s_t$ the signal at time $t$.

Finally, we performed "leaky" integration of the derivative by calculating the exponentially-weighted moving average (EWMA) with half-life $\tau_{half} = 0.4$ s (i.e., time of half-maximum). Note that this operation is equivalent to convolution with an exponential kernel. The operation that yields the filtered time series $y_t$ from the low-passed sensor signal $x_t$ can be expressed as in equation 3:

$$y_t = (1-\alpha) \cdot y_{t-1} + \alpha(x_t - x_{t-1}), \tag{3}$$

with

$$\alpha = 1 - e^{\log\left(\frac{1}{2 \cdot \tau_{\text{half}} \cdot \Delta t}\right)}, \quad (4)$$

where $\Delta t$ refers to the length of the time step used.

On this filtered signal we searched for "bouts" of rising amplitude. We first computed the differential of $y_t$ as in eq. 5:

$$y'_t = y_t - y_{t-1}. \quad (5)$$

A bout is then characterised by $y'_t$ being equal to or larger than zero. We can define the variable $b_t$ that is 1 if a bout is present at time $t$, and 0 if no bout is present (eq. 6)

$$\begin{array}{ll} b_t = 1 & \text{if } y'_t \geq 0 \\ b_t = 0 & \text{otherwise} \end{array}. \quad (6)$$

A bout onset is detected if $b_t$ flips from 0 to 1, and the bout lasts until $b_t$ flips back to 0 again. The bout amplitude $a_{\text{bout}}$ is defined as

$$a_{\text{bout}} = y_{t_2} - y_{t_1}, \quad (7)$$

with $t_1$ and $t_2$ the start and end time of the bout. Note that since $y_t$ is monotonously rising between $t_1$ and $t_2$ (as $y'_t \geq 0$, eq. 6), $y_{t_1}$ and $y_{t_2}$ correspond to the minimum and maximum of the signal within the bout.

Putative false-positive bouts were filtered out using an amplitude threshold. Assuming that gas-induced bouts occur only during gas release, we consider bouts detected before gas release started as false-positive. We estimated the amplitude threshold using a three-sigma criterion,

$$\theta_{\text{amp}} = \langle a_{\text{blank}} \rangle + 3 \cdot \sigma_{\text{blank}}, \quad (8)$$

with $\langle a_{\text{blank}} \rangle$ the average amplitude and $\sigma_{\text{blank}}$ the standard deviation of the amplitude of events during the blank period (usually until $t = 50\ s$), in all trials that were available for the specific gas/sensor/board combination (usually $n = 20$).

## 2.3  Software and reproducibility

We used the `pandas` package to parse, clean and resample the original data files. Spectral decomposition and filtering was performed using the `numpy.fft` ([numpy.org](numpy.org), [15]) and `scipy.signal` packages ([scipy.org](scipy.org), [16]). All figures were prepared from raw data using the `matplotlib` package ([matplotlib.org](matplotlib.org), [17]). The code for creating the figures was written in IPython/Jupyter notebooks ([ipython.org](ipython.org), [18]). We used the packages within the Anaconda python distribution (Continuum Analytics, Austin, TX, USA, [continuum.io](continuum.io)).

All analysis code used in this study is freely available under an open source license at https://github.com/BioMachineLearning/exploiting_plume_structure. This repository also contains IPython Notebooks which allow the reader to completely reproduce the entire analysis, including recreation of all figures in this study.

## 3 Results

### 3.1 Spectral analysis of e-nose recordings

We analysed signals from electronic nose sensors recorded in a wind tunnel at varying distance from the gas source. With increasing distance from the source, the amplitude of the signal decreases, but the fluctuations also became less rapid (Fig. 2). This observation suggested that the signal contained less high-frequency components when it was recorded further away from the source.

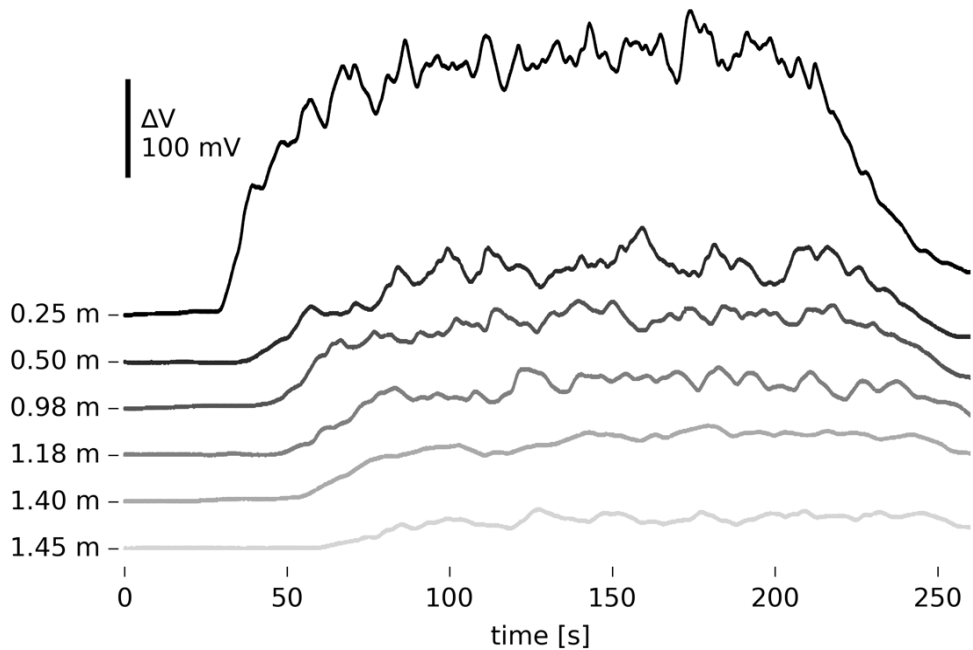

**Fig. 2:** Responses of a TGS2610 gas sensor to Acetaldehyde (500 ppm), in a wind-tunnel, at varying distance from the gas. While the signal amplitude clearly encodes the distance to the source (high gas concentration close to the source), the speed of the fluctuations of the signal apparently correlates to distance as well.

To test this assumption, we extracted the frequency band that contains those fast fluctuations. We split the signal into three bands using a band-pass filter (2nd order Butterworth, critical frequencies: 0.025 Hz, 0.5 Hz). The low-frequency domain contained the constant offset and slowly varying parts of the signal that originated from the onset and offset of gas delivery (Fig. 3A). The medium-frequency domain encoded small, rapid fluctuations of the gas concentration, potentially due to the filamentous nature of gas distribution in the turbulent environment of the wind tunnel (Fig. 3B). The high-pass band mainly contained electronic noise of small amplitude (Fig. 3C). We further analysed the middle band.

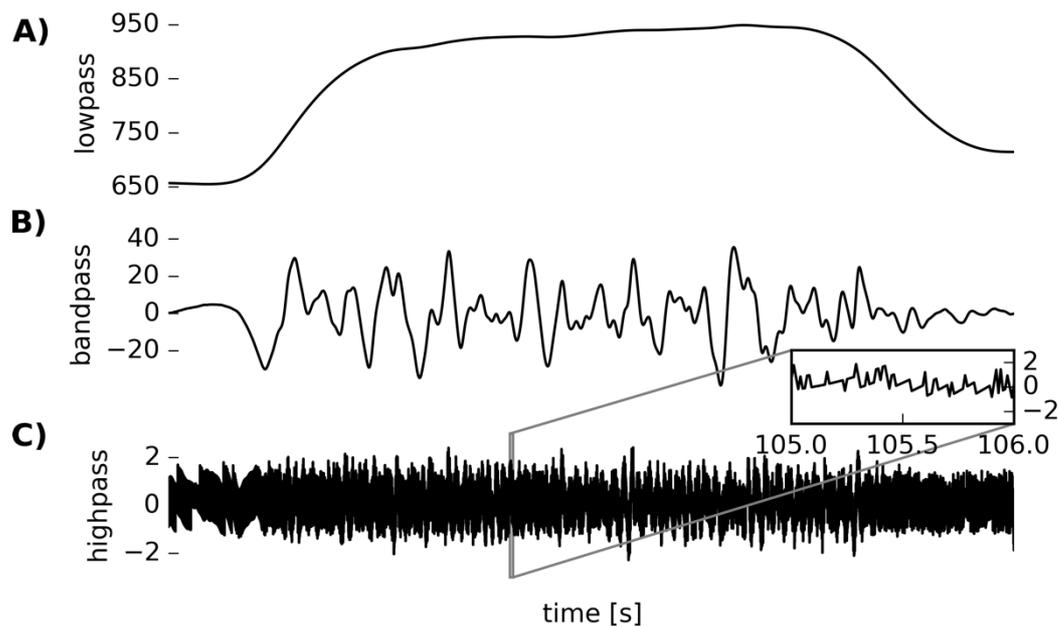

**Fig. 3:** Splitting the signal recorded at $d = 0.25\,m$ into low, middle, and high frequency components (critical frequencies: 0.025 Hz, 0.5 Hz). **A)** The low frequency component follows the onset/offset time scale. **B)** The band-passed signal carries information about fast concentration fluctuations. **C)** The high-passed signal contains mainly noise.

We observed that fluctuations in this band vary more rapidly in those recordings that were taken close to the source, and slower at a larger distance (Fig. 4A). This observation is also visible in the spectral decomposition of the band-passed signal (Fig. 4B): The recordings closer to the source contain more power in higher frequencies relative to the lower frequency band. In order to quantify the relative amount of power in high frequencies, we devised the estimator $P_{\text{rel}}$ as the fraction of power above a cut-off frequency $f_{\text{crit}}$ (eq. 9):

$$P_{\text{rel}} = \frac{\sum_{f_{\text{crit}}}^{f_{\text{nyquist}}} p_f}{\sum_{0}^{f_{\text{crit}}} p_f}, \qquad (9)$$

with $f_{\text{nyquist}}$ the Nyquist frequency of the signal. The cutoff frequency $f_{\text{crit}}$ was estimated as the average maximum frequency of the six spectra (corresponding to the six distances). By linear regression we could identify a relationship between $P_{\text{rel}}$ and source distance (Fig. 4C). This relationship could be used to provide a rough prediction of source distance from the spectral feature $P_{\text{rel}}$ with an average RMSE of $0.30\ m \pm 0.06\ m$ (five-fold cross-validation over all 20 trials). Thus, we concluded that source distance is encoded in fast fluctuations of gas concentration, and that off-the-shelf MOX gas sensors provide sufficient temporal resolution to exploit this information for distance prediction.

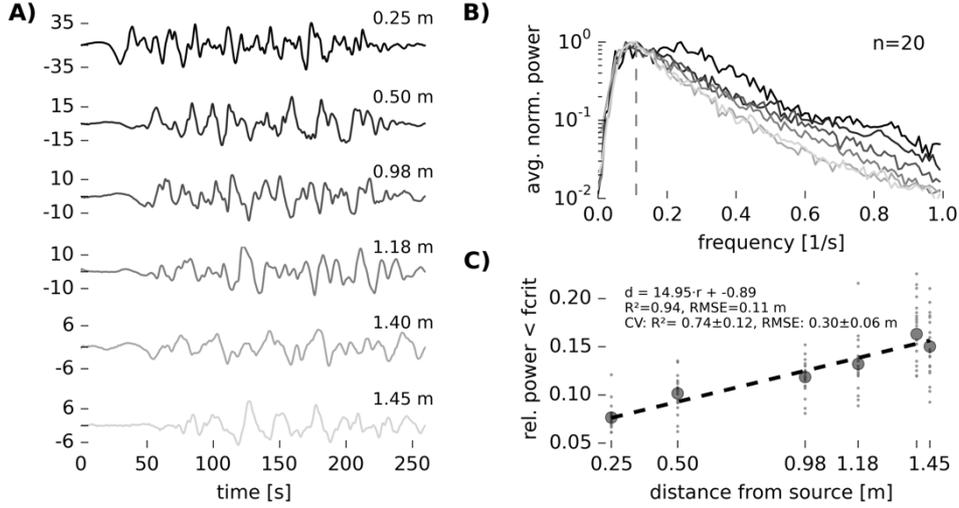

**Fig. 4:** Spectral analysis of band-passed signal. **A)** Band-passed signal as acquired at various distances from the gas source. **B)** Average power spectra of band-passed signals at all distances. Shades of grey encode source distance as in panel A. Average of $n = 20$ trials at each location. Spectra were normalised to unit maximum power. Dashed line: $f_{crit}^{low\ vs. high}$, cut-off frequency for relative power comparison. **C)** Fraction of total power below $f_{crit}^{low\ vs. high}$ vs. sensor location. The farther the sensor, the more power is contained in the part of the spectrum below $f_{crit}^{low\ vs. high}$ (eq. 9). Large grey dots: average from 20 trials, small dots: individual trials. Dashed line: regression on the average relative power values against distance from source.

The potential cause for this effect lies in the fact that the number of plume filaments detected in a fixed time interval correlates with the distance to the source [6] (which has been shown using photo-ionisation detectors which yield a much higher temporal resolution than MOX-sensors). We next analysed whether we could identify encounters of individual plume filaments ("bouts") using MOX-sensors.

## 3.2 Decomposition into "bouts" of gas detection

The impulse response of a MOX sensor is thought to be characterised by a fast initial response, which is however dominated by a slow recovery, caused by the slow reversal of the reaction of the volatile with the sensing electrode. Hence, when a plume filament hits a sensor, the signal will rise quickly and taper off slowly afterwards. The slow decay of the signal tends to overshadow the fast rising flanks of the subsequent responses, leading to the perception that responses of MOX-sensors are "slow", when in fact, they are mostly only slow in recovery. To overcome this limitation, we adopted a filtering approach that was previously used to accelerate sensor calibration [12], to evaluate the transient phase of a MOX-sensor signal for odour recognition [13], and for enhanced change-point detection in continuous gas stimulation [14]. Briefly, we first applied a low-pass filter to attenuate high-frequency electronic noise. Under the assumption that the initial reaction of the analyte with the electrode is very fast compared to its reversal, we eliminated the slow component by forming the differential. The resulting signal is subsequently convolved with an exponential kernel (i.e., a "leaky integrator", eq. 3).

Compared to band-passing as in Fig. 3, the proposed filter was more efficient in removing slow changes Fig. 5A. Besides, the operation is computationally simpler than Butterworth filtering: both differentiation and leaky integration are causal operations, i.e. they only depend on precedent values of the signal (see eq. 3), and only the value obtained in the previous time step must be stored. The procedure thus lends itself to a straightforward implementation with minimal resources for computing and memory, e.g. on a low-spec microcontroller.

Our next aim was to detect gas plume encounters, i.e., to identify portions of the signal during which it was continuously rising. To this end, we isolated portions where the derivative of the filtered signal is positive (eq. 5 and 6). Fig. 5A highlights these portions of the signal, which in the following we will refer to as "bouts".

## 3.3 Distance prediction with bouts

Previous studies with photo-ionisation detectors indicated that the number of plume filament encounters decreases with increasing distance from the source [2,6]. We therefore analysed how the number of bouts detected in the response from MOX-sensors relates to their distance from the gas source. We found that the number of bouts is indeed a strong predictor for source distance, with fewer bouts being encountered as the distance increases (Fig. 5B). Note that the measurements at $d = 1.45\ m$ present an outlier in that the bout count increased slightly compared to $d = 1.40\ m$. This may be due to the fact that the measurements at $d$ = 1.45 m have been acquired very close to the outlet of the wind tunnel, and the structure of the turbulent gas plume was likely distorted by the imperfect aerodynamics in this area (this circumstance is also discussed in the original reference [7]).

We fitted a linear model relating bout counts to source distance and assessed its performance in cross-validation (5-fold) over 20 trials. The resulting models were able to predict source distance with an average $RMSE_{CV}$ of $0.18 \pm 0.03\ m$ (mean ± 95% confidence interval). Most notably, the bout count feature predicted source distance with considerably higher precision and lower error than the spectral feature used in Fig. 4 ($RMSE_{CV} = 0.74 \pm 0.12\ m$). We wish to note that while it is unlikely that bout counts are linearly related to source distance, a linear fit is the most robust when the true relationship is not known (but see section 3.7 below for a detailed analysis of the shape of the bout count-to-distance curve).

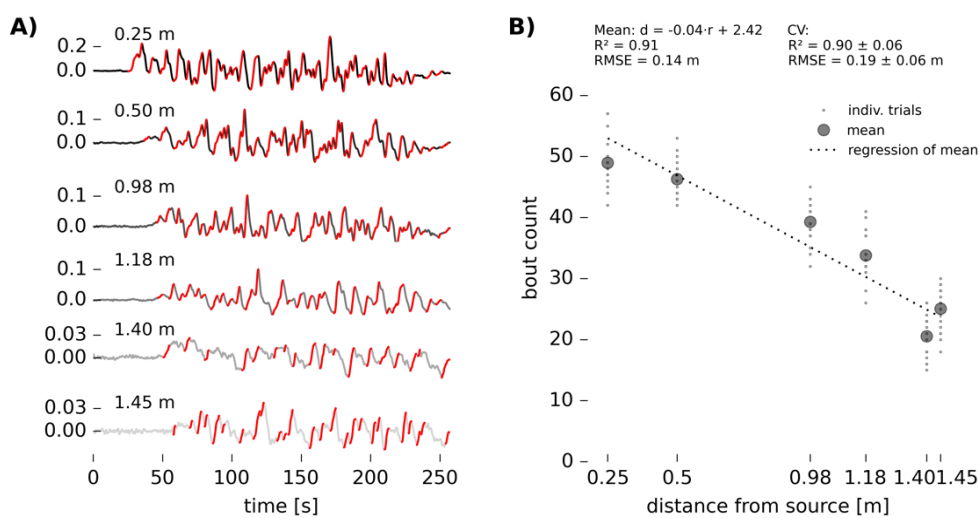

**Fig. 5:** Bout-based distance prediction. Acetaldehyde 500 ppm, sensor TGS2610, trial number 10 (same data as in Fig. 2). **A)** MOX-sensor signals after applying the three-step low-pass/derivative/EWMA-filter. "Bouts" in the signal are coloured red. **B)** Bout counts for 20 trials vs. distance from source. Numbers under "mean" refer to a regression of the mean counts. R2 and RMSE for cross-validated regression (5-fold) are given under "CV".

We found that distance prediction is also possible at different wind speeds, although we found that for optimal results the wind speed should make part of the regression variables (see supplemental Fig. S2 and accompanying text). Moreover, we analysed the influence of the sensor voltage (i.e., the heater temperature) on distance prediction, and found that best results were achieved with heater voltage between 5V and 6V (see supplemental Fig. S3 and accompanying text).

### 3.4 Sensor invariance of signal bouts

The examples above were obtained using one sensor, but we found that the detection of bouts is possible on all sensors that responded to the gas under scrutiny (Fig. 6). Even for those sensors where the raw signals show hardly any fast fluctuations dependent on gas concentration (e.g. sensors TGS 2602, 2600a, and 2600b in Fig. 6A), the filtering procedure revealed the bouts (Fig. 6B).

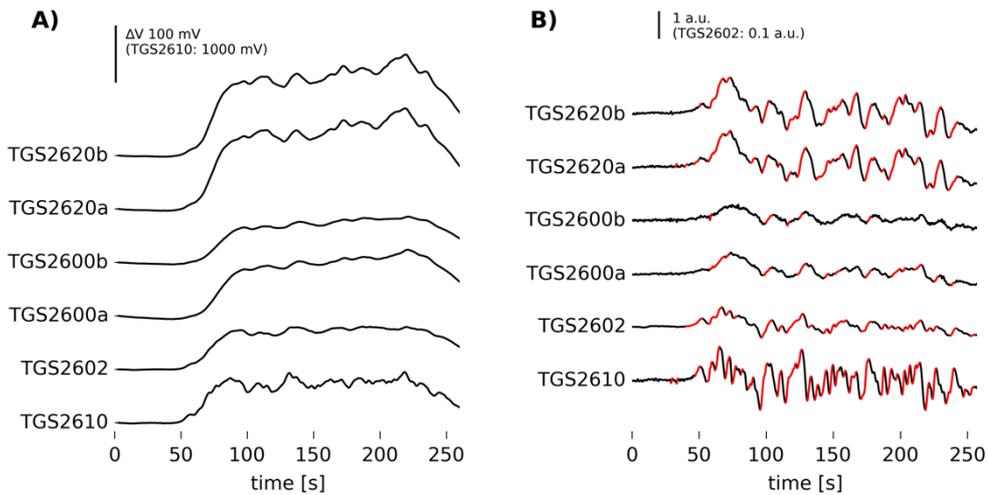

**Fig. 6:** Bout analysis of all sensors on one sensor board, $d = 1.18\ m$, trial 19. **A)** Raw sensor response. Sensor 3 is scaled down by factor 10. Sensor 2 did not respond to the gas, sensor 1 was excluded because it contained sampling artefacts. **B)** Filtered response with detected bouts highlighted in red (*cf.* Fig. 5A).

On all sensors that showed a response, these bouts occurred at highly correlated time points. This was particularly visible in pairs of identical sensors, like the two instances of TGS 2620, for which the onset of the bouts coincided and were of practically identical shape (Fig. 7A). But also on sensors with markedly different tuning, bout onset times were strongly correlated, e.g. for TGS 2610 and 2602 (Fig. 7B). We assessed the correlation between onset times across all sensors using an approach from [19] (see supplemental figures S4 and S5 and accompanying text), and found it to be significantly higher than random. This observation suggests that bouts are likely caused by a common external source, like encounters of high-concentration filament of the gas plume.

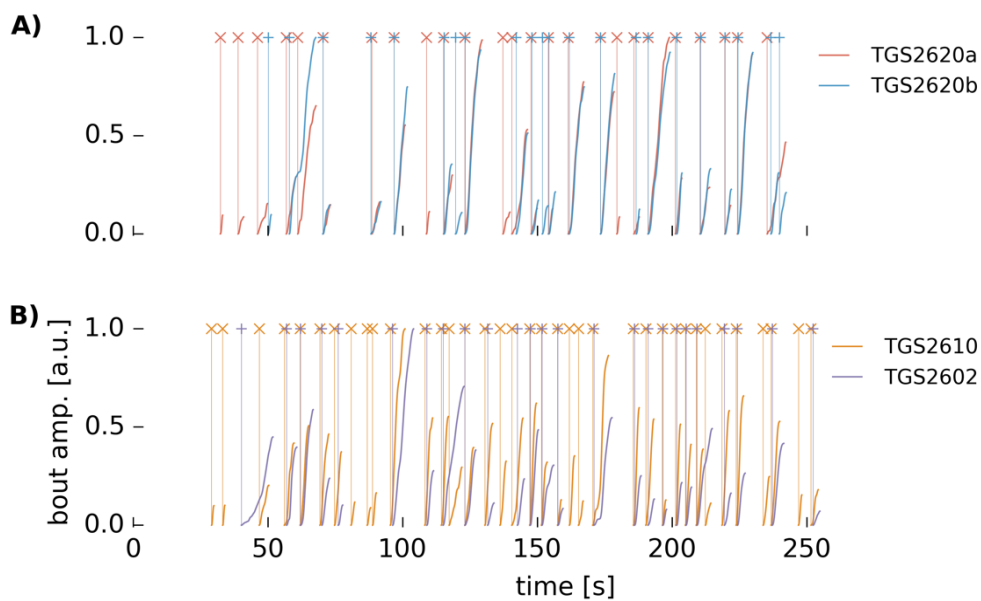

**Fig. 7:** Coincidence of bouts on A) two sensor of the same model, B) two different sensor models.

### 3.5 Gas dependence of bout-based distance prediction

Next, we asked whether the relationship between bout frequency and source distance can also be established for other gases, using the same procedure for cross-validated distance prediction as in Fig. 5. Since not all gases were equally well resolved by the one single sensor, we show the performance of cross-validated distance prediction for all gases in the data set, using the the sensor that yielded the smallest RMSE$_{CV}$ (Fig. 8A-K). The distance prediction was possible for all gases using single sensors, with an average RMSE of 0.17 $m$.

One gas, Carbon Monoxide (CO), was presented at two concentrations – 1000 $ppm$ (Fig. 8F) and 4000 $ppm$ (Fig. 8G). Distance prediction worked for both concentrations, but the lower concentration yielded consistently fewer bout counts. In general, since bouts are a consequence of advective processes in a turbulent environment, we expect the bout count feature to be largely concentration invariant. However, since the lower concentration also comes with a lower signal-to-noise ratio, some proportion of bouts may have been falsely rejected as noise due to their small amplitude, leading to seemingly lower bout counts.

While a strong relationship between the number of detected bouts and the distance to the source is discernible for all gases, there is a comparably large variation in the absolute bout counts across gases at each position. Moreover, the slopes of the fitted curves in Fig. 8A-K differ considerably between gas species. There is a weak positive correlation between the slope of the regression line and the molecular weight of the gas (Fig. 8L, correlation coefficient $cc_{\text{pearson}} = 0.54$, $p = 0.09$). Hence, the number of detected bouts falls more steeply with distance for lighter gases. This observation may be related to the buoyancy of gases, as plumes of lighter gases may tend to rise as they drift through the tunnel, and potentially interact less with the sensors that are mounted on the bottom.

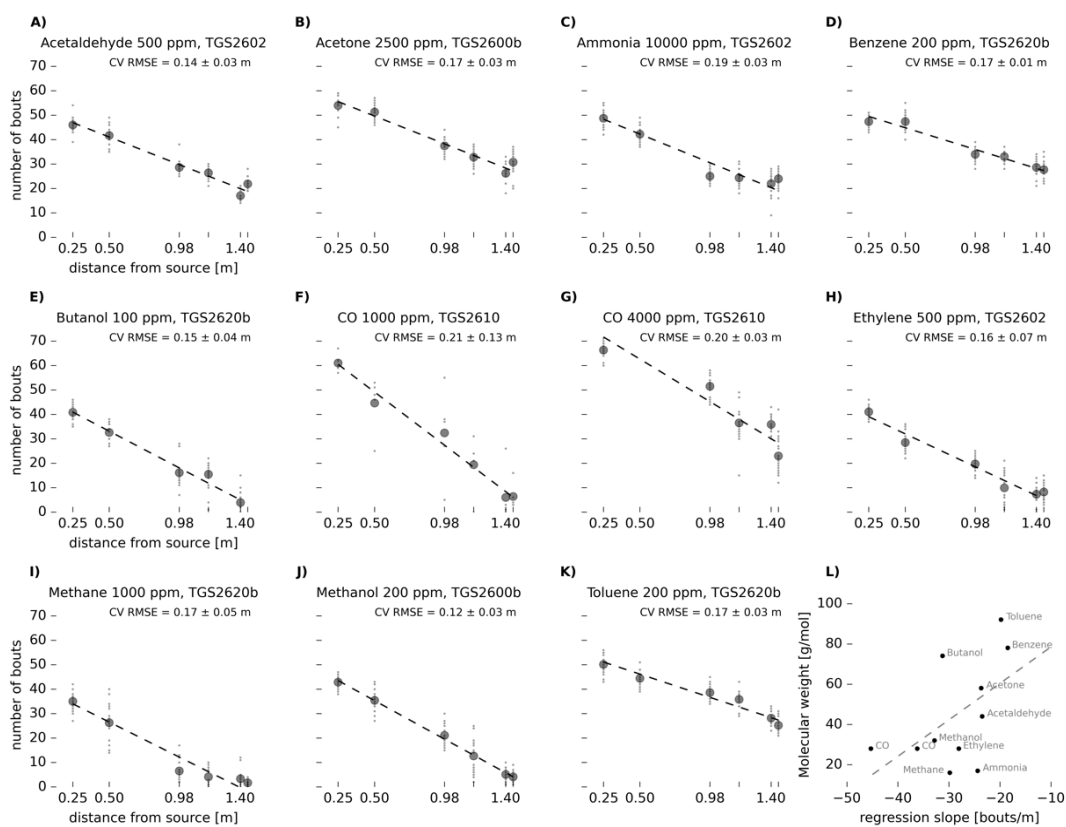

**Fig. 8:** Relationship of bout counts to distance from source for all gases. **A-K)** Bout count vs. distance for each gas. Dashed line: linear regression of the mean bout count over 20 trials. RMSE of 5-fold cross-validation (CV) ± 95% confidence interval is given in the upper right of the panel. For each gas the sensor was picked that yielded the best RMSE in CV. **L)** Slope of regression vs. molecular weight, correlation coefficient $0.54, p = 0.09$.

Since most sensors are capable of resolving bouts for several gases, we next explored whether multivariate regression with bout counts from all sensors would improve distance prediction. So instead of training the regression model separately for each gas and sensor and selecting the best model, we used the 7-dimensional vector of bout counts from all sensors (except sensor 1), and test its performance in cross-validation. For 7 out of 11 gas species, the RMSE obtained with multivariate regression was lower than the best single-sensor RMSE (see Fig. 9). For the remaining 4 species, the multivariate regression still performed well within two standard deviations of the cross-validation RMSE. Thus, on average, using bout counts from several receptors improves distance prediction.

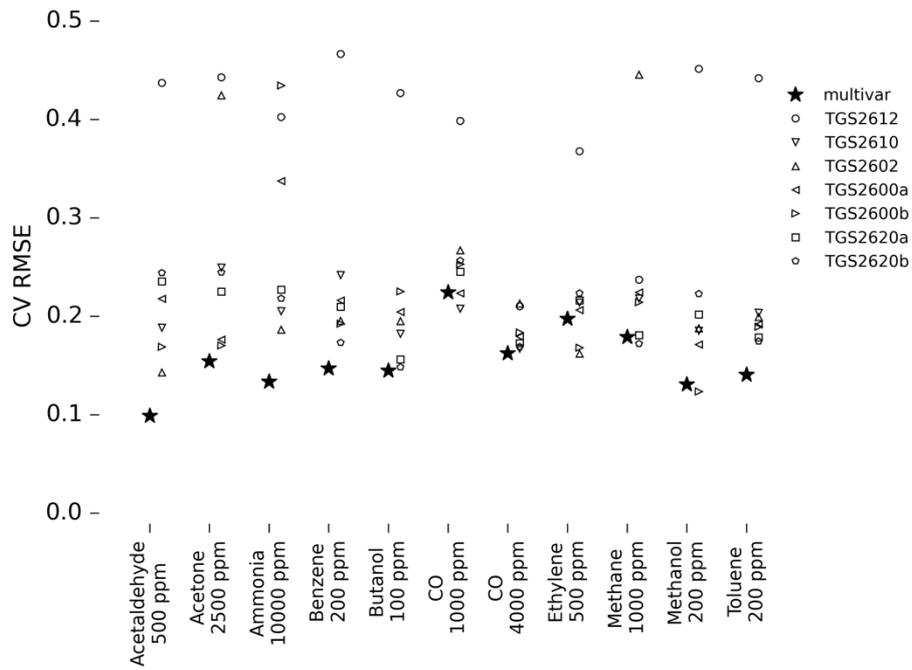

**Fig. 9:** Multivariate vs. univariate regression.

### 3.6 Cross-wind statistics of bout counts

So far, we analysed bout count statistics only straight downwind from the plume (i.e., along the centre of the wind tunnel, using board no. 5). We next investigated how bout statistics vary in a cross-wind fashion, i.e., perpendicular to the downwind axis. At the measurement position closest to the source ($d = 0.25\ m$), bouts were detected almost exclusively at the central sensor board, no. 5 (Fig. 10A). At larger distances, considerable numbers of bouts were detected at positions lateral to the centreline, first at $d = 0.5\ m$ on board no. 4, then at $d = 0.98$ m on board no. 6. When taking the location of first bout encounter as a reference, bout counts gradually decreased with increasing distance from the source. This suggests that average bout count is a viable proxy for source distance also when the measurement device is not located directly downwind from the source, as long as it is located within the region of gas plume dispersal.

Only small numbers of bouts were detected on the three outermost boards on either side (no. 1, 2, 3 and 7, 8, 9). This observation indicates that the region of plume dispersal was mainly limited to a region central to the wind tunnel (see also [7]).

Interestingly, the standard deviation of average bout counts across repetitions was low along the centreline, even for large distances (Fig. 10B). In contrast, the variance of the bout counts was considerably higher for detectors located away from the centreline.

This result is consistent with previous observations in the open field [2,4] and in wind tunnels [5,6], according to which the intermittency of gas encounters increases when moving farther away from the centreline of the plume. These findings suggest that the variance of bout counts could be used as an indicator as to whether the detector is located directly downwind of the source, or rather slightly lateral from it.

Notably, the function $std(n_{\text{bout}})/\langle n_{\text{bout}}\rangle$, i.e., the standard deviation of bout counts normalised for the average number of bouts encountered, has an absolute minimum at the centreline (Fig. 10C). This would allow a mobile agent to locate the centreline of the plume by trying to minimize bout count variance in its search space.

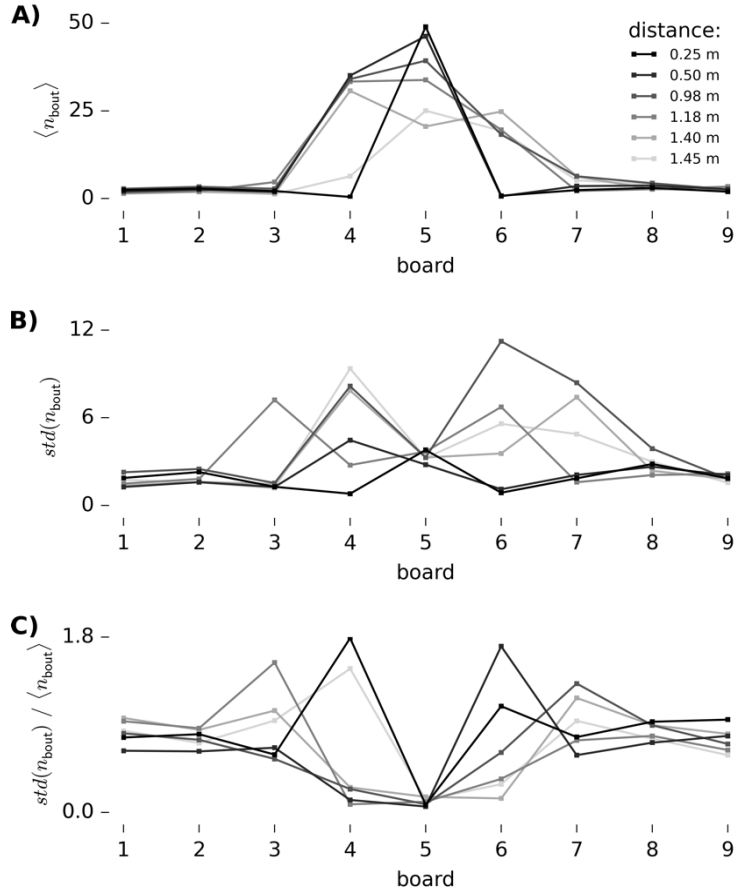

**Fig. 10:** Cross-wind bout count statistics; Acetaldehyde 500 ppm, heater voltage 6 V, fan speed 1500 rpm. **A)** Average bout counts $\langle n_{\text{bout}} \rangle$ over 20 repetitions. Board no. 5 is the central board. Distances are color-coded as indicated in the legend. **B)** Standard deviation of bout counts $std(n_{\text{bout}})$ over 20 repetitions. **C)** Standard deviation of bout counts over their average $std(n_{\text{bout}})/\langle n_{\text{bout}} \rangle$.

## 3.7 Influence of kernel parameters on bout counts

The analysis we present here has two free parameters: The width of the Gaussian kernel for initial low-pass filtering of the signal, $\sigma_{\text{smooth}} = 0.3\ s$, and the halflife of the exponential kernel $\tau_{\text{half}} = 0.4\ s$ for "leaky" integration of the derivative of the signal. We initially tuned these parameters manually to yield good results with data set at hand. But the choice of parameter values certainly influences the result of the filtering process, and hence the detection of gas bouts. Therefore, we next explored the effect of those parameters.

Our filtering approach can be expressed in terms of subsequent, discrete convolution of the signal $S$ by three kernels: A Gaussian kernel $G$ with width $\sigma_{smooth}$, a differentiating kernel $D$ which is of the form [1, -1] and an exponential kernel $E$ with halflife $\tau_{half}$. Hence, the filtering process can be expressed as in eq. 10,

$$S_{filt} = S * G * D * E, \qquad (10)$$

with the asterisk * denoting the convolution operator. Since convolution is associative, eq. 10 is equivalent to eq. 11,

$$S_{filt} = S * F, \quad \text{with} \quad F = G * D * E. \qquad (11)$$

The two parameters $\sigma_{smooth}$ and $\tau_{half}$ govern the shape of the kernel $F$ (Fig. 11A and C). Higher values $\sigma_{smooth}$ and $\tau_{half}$ lead to a more elongated kernel shape (Fig. 11B and D). The frequency response of F corresponds to a band-pass filter, and more elongated kernels cause the frequency response of $F$ to be shifted towards lower frequencies.

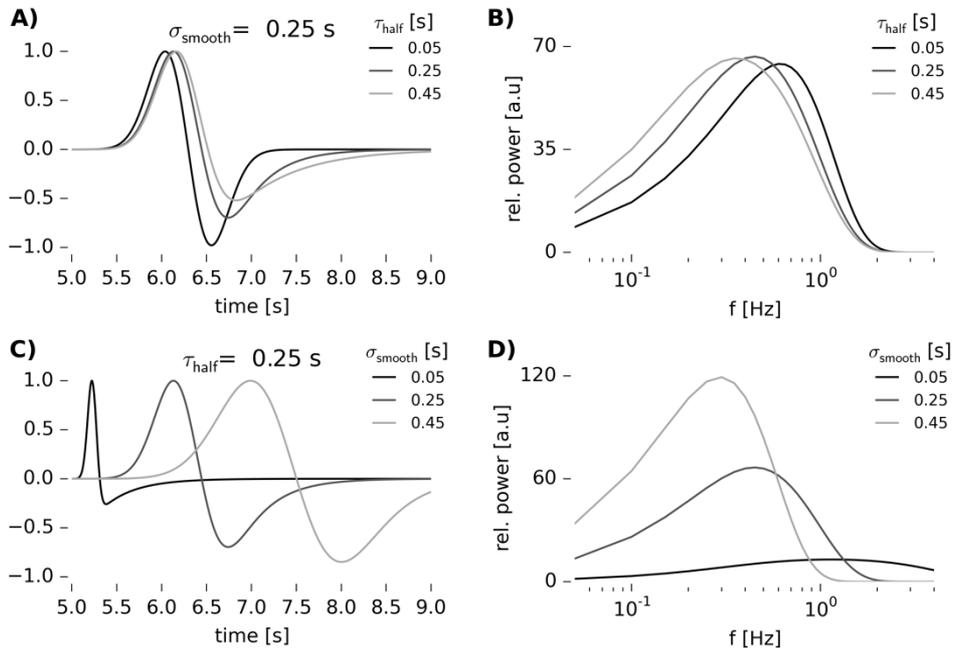

**Fig. 11:** Parameter influence on the shape of the filter kernel $F$ and its frequency response. **A)** Impulse response of the filter $F$ when varying $\tau_{half}$ while keeping $\sigma_{smooth} = 0.25\ s$ constant. **B)** Frequency response of $F$ associated to the kernels shown in A). **C)** Impulse response when varying $\sigma_{smooth}$ while keeping $\tau_{half} = 0.25\ s$ constant. **D)** Frequency response of the filter in C).

Next, we analysed how $\tau_{half}$ and $\sigma_{smooth}$ influenced distance-dependent bout counts, uncovering rather complex relationships. For small values of $\sigma_{smooth}$ and $\tau_{half}$, the shape of the bout count curve was convex, similar to an exponential decay, with higher bout counts detected close to the source, and few bouts at greater distance (Fig. 12). Conversely, larger values of $\tau_{half}$ and $\sigma_{smooth}$ yielded fewer bout counts close to the source, but with a weaker decay of bout counts towards greater distances. Notably, either keeping $\sigma_{smooth}$ constant and varying $\tau_{half}$ (Fig. 12A) or the other way round (Fig. 12B) had the same effect on the shape of the curve: it gradually switched from convex to concave.

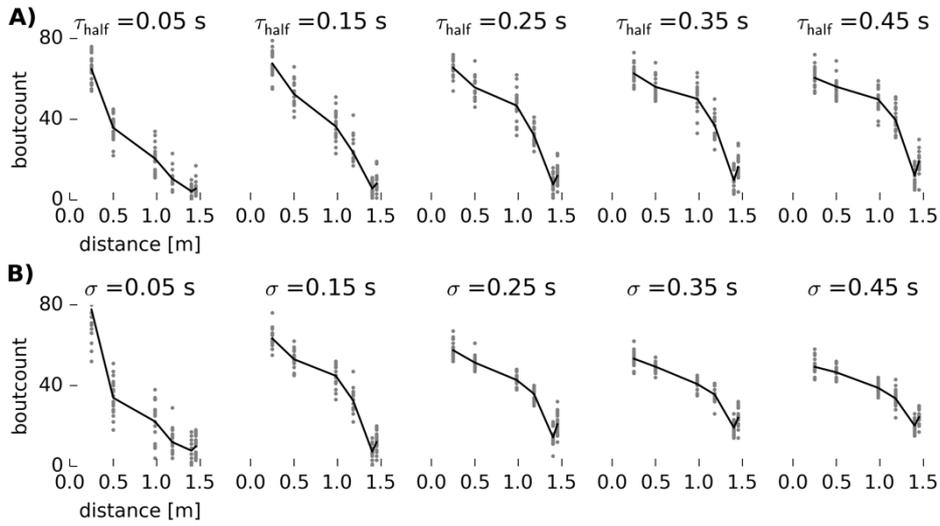

**Fig. 12:** Effect of filter parameters on bout count curves. **A)** Varying $\tau_{half}$ while keeping $\sigma_{smooth} = 0.1$ s constant. **B)** Varying $\sigma_{smooth}$ while keeping $\tau_{half} = 0.15\ s$ constant.. In both cases. the curve shape shifts from concave to convex.

The shape of these curves can be explained by the expected number of filaments in a model of plume growth, in combination with the band-pass characteristics of the filters we employed. Close to the source, the individual filaments of the gas plume are comparably small in diameter, likely in the range of the diameter of the gas outlet. Further away from the source, the extent of the filaments increases by diffusion and advective processes. The bout length, i.e., the average amount of time that a sensor will be within such a filament will hence be shorter close to the source (although long bouts will still occur, e.g. when a filament traverses the sensor longitudinally). Advective processes move the plume and stretch the filaments as source distance increases, leading to longer bouts. The short bouts encountered close to the source will be better resolved by filters tuned to higher frequencies (i.e., smaller $\tau_{half}$ and $\sigma_{smooth}$), but perform less well in detecting longer bouts at larger distances. Likewise, filters tuned to lower frequencies may fail to resolve many of the sharp bouts in close source proximity, but instead perform better on the long bouts that occur further away from the source.

We wish to emphasise that any filter setting that we tested yielded the same general relationship between bout count and source distance. However, the fact that the shape of the bout-count-to-distance curve is strongly influenced by parameters of the filter indicates that a general assumption about the nature of the non-linear relationship between bout counts and source distance cannot be made with the present data. Hence, our approach to predict source distance with a linear model appears well justified, even though the true relationship is likely non-linear. The linear fit thus provides a lower bound for the accuracy of the estimation of source distance.

A potential route to extract more information from the signals and to increase the accuracy of the distance estimate may lie in using several different filter settings in parallel. Our observations in Fig. 12 indicate that the filter could be tuned to provide particularly good resolution at a certain distance range, e.g. by adjusting $\tau_{half}$ and $\sigma_{smooth}$ such that the slope of the bout-count-to-distance curve is maximal in the region of interest. Smaller values of $\tau_{half}$ and $\sigma_{smooth}$ may be preferred if the expected distance is short, while larger values may be better suited to resolve longer distances. A "filter bank" that implements a range of settings for $\tau_{half}$ and $\sigma_{smooth}$ could be designed to cover different parts of the frequency spectrum and cover larger distance ranges in parallel. Such an approach may become particularly relevant in more practical scenarios, such as larger wind tunnels or, ultimately, in open field conditions.

### 3.8 Comparison of the filtering approach with de-convolution

As stated above, it can be assumed that the impulse response of a MOX-sensor is characterised by a fast onset and a slow decay. The time-dependent impulse response $r(t)$ of a MOX-sensor can thus be approximated by a bi-exponential function of the form (eq. 12),

$$r(t) = \frac{-b}{b-a} \cdot \left( e^{-\frac{t}{a}} - e^{-\frac{t}{b}} \right), \tag{12}$$

with $a$ and $b$ controlling the extent of the decay- and rise-time, respectively.

Assuming that an impulse response of this shape is sufficiently close to the true impulse response, it becomes possible to "de-filter" the sensor response to retrieve the original time course of the gas concentration. De-filtering can conveniently be achieved by division in frequency space, i.e., de-convolution, according to eq. 13,

$$\hat{\mathbf{s}}_{\text{deconv}} = \hat{\mathbf{s}} / \hat{\mathbf{r}}, \tag{13}$$

with $\hat{\mathbf{s}}$ and $\hat{\mathbf{r}}$ the representation of the signal and the impulse response in the frequency domain (as obtained by discrete Fast Fourier Transform, FFT), and $\hat{\mathbf{s}}_{\text{deconv}}$ the de-convolved signal in frequency domain. The de-convolved signal in the time domain can be recovered from $\hat{\mathbf{s}}_{\text{deconv}}$ by inverse FFT. Before transforming the signal to the frequency domain we removed high-frequency noise and constant offset by filtering it using a band-pass filter with critical frequencies 0.0005 Hz and 0.5 Hz. We marked portions of the signal that had positive slope as "bouts".

We estimated the parameters of the bi-exponential model for the sensor's impulse response according to two criteria. First, since gas concentration cannot be negative, the de-convolved signal should not fall below zero at any point during the measurement. Second, assuming that the gas concentration dropped to non-detectable levels after the gas release stopped at around $t = 200\ s$, we postulated that the de-convolved signal should approach zero towards the end of the recording. Following these guidelines, we manually estimated the parameters of the impulse response to be $a = 0.1\ s$ and $b = 10\ s$.

Fig. 13A shows the de-convolved signal together with the raw signal, and Fig. 13B shows the kernel used for de-convolution. Fig. 13A also shows the detected bouts, marked in red. For comparison, we also show the same signal treated with the filtering and bout detection approach outlined in the previous sections (Fig.

13C). The bout structure is virtually identical between the de-convolved signal and the filtered signal (see close-up in Fig. 13D).

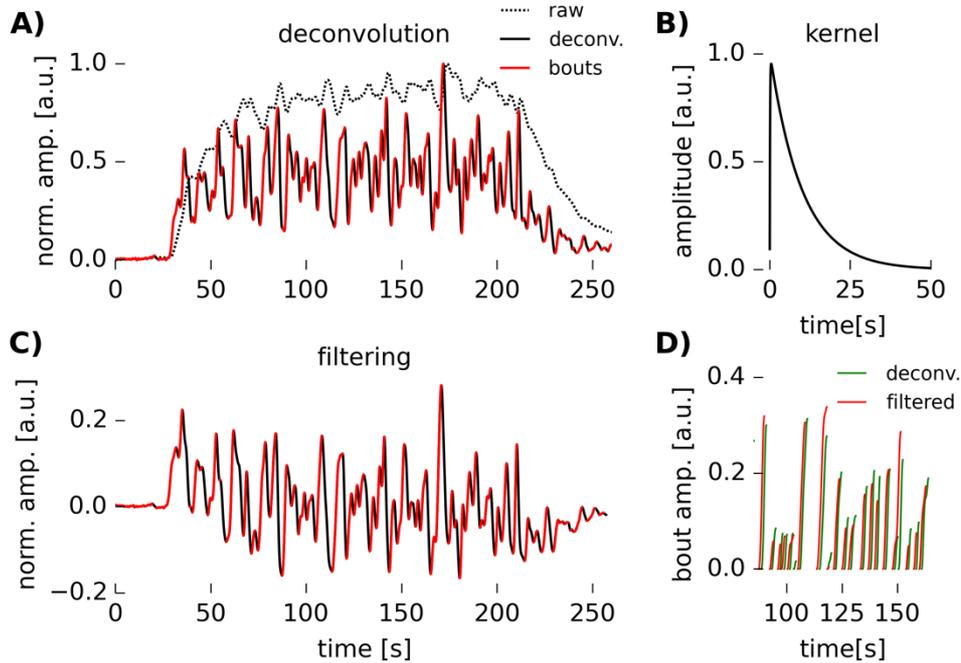

**Fig. 13:** Bout detection by de-convolution. **A)** The raw signal and the signal after de-convolution with a bi-exponential kernel. The detected bouts are marked in red. **B)** The bi-exponential kernel used for de-convolution. **C)** For comparison, the same signal as in A), after filtering and bout detection as outlined above. **D)** Close-up on the bouts detected by either method in the interval between 85 and 165 s.

This observation indicates that if the primary goal of the signal analysis is the detection of bouts, the cascaded filtering approach outlined in the previous sections is equivalent to model-based de-convolution of the signal. De-convolution yields an accurate estimation of the time-dependent gas concentration at the sensor's location (if the estimated impulse response is sufficiently close to the real impulse response). The advantage of the cascaded filter model is that is computationally considerably less complex, not only because of the arithmetics involved in FFT, but also in terms of memory requirements. Moreover, for meaningful results, FFT must be applied on the whole signal, or at least on a temporal segment of significant size (as in a windowed approach). In contrast, the cascaded filtering can be implemented in a fully streaming mode that does not require windowing. Taken together, these observations emphasise the utility of the proposed approach on low-power, embedded hardware that might be employed in energy-restricted scenarios, like e.g. mobile robotics.

## 4 Discussion

We have shown that it is possible, using MOX sensors, to decode temporal features of odour plumes that encode source distance in a turbulent environment. To this end, it was necessary to filter the signal using a cascaded filtering approach: After removing high-frequency noise, we first form the differential of the signal and subsequently perform a leaky integration using an exponential kernel. Then we identify "bouts" in the filtered signal, i.e., portions with upward slope, that may correspond to the sensor encountering plume filaments with high gas concentration. The number of such bouts decreases as the distance to the source increases. Since the "bout count" feature is largely independent of gas concentration, it complements concentration-based approaches to estimate source distance in turbulent environments.

While this observation is in line with findings obtained with PIDs [2,4–6], we wish to stress that this study is the first to demonstrate that this effect can in fact be exploited with low-cost, off-the-shelf MOX sensors. The fact that the statistics of concentration fluctuation can be decoded from MOX sensors, in spite the time constants of their gas response being orders of magnitude longer than PIDs, may be a consequence of the self-similar organisation of turbulent plumes on the spatio-temporal scale. On the other hand, the comparably long sensor response time constants also entail longer measurements (compared to PIDs) until a robust estimate of source distance can be obtained. However, recent findings on speeding up MOX sensor constants by modulating the temperature of the heater element may allow to increase the temporal resolution of MOX sensors, and accelerate the decoding of distance-dependent fluctuations in gas concentration [20].

The signal processing algorithm that we used requires only minimal resources in terms of memory and computational capabilities. It is straightforward to implement on low-power microcontrollers, and could easily deployed in environments that impose strong restrictions on power consumption, like mobile robotics. Indeed, a study exploring methods to estimate the proximity of a gas source using MOX sensors in a turbulent environment found that the variance of the gas concentration can be better indicator for source distance than the time-averaged sensor response [3]. The analysis we present here extends the concept of variance detection by detecting events of high gas concentration, which significantly increased the accuracy of source distance estimation. Combining the proposed bout count feature with robotic navigation therefore may present an interesting remit for future studies.

Finally, we described how the two parameters of the signal processing method can be used to tune the method to certain ranges of gas intermittency. If the filter is tuned to higher frequencies, it will resolve distance best if it's close to the source. Larger distances will be better resolved when tuning the filter to lower frequencies. Since the simple design of the filter allows to use several, differently tuned instances simultaneously, one could construct a "filter bank" that resolves a wide range of different source distances. Interestingly, such a filter bank tuned to varying degrees of gas intermittency has been described in arthropod olfaction [21].

## 4.1 A putative strategy for gas-based robot navigation using bout counts

We have observed two features of gas plumes that carry information about the location of the detector relative to the source. First, the number of gas bouts gives rise to an estimate about the distance to the source. Second, the variance of those bout counts over time provides information about whether the detector is located straight downwind (low variance) or slightly lateral from it (high variance). This is summarised in Fig. 14.

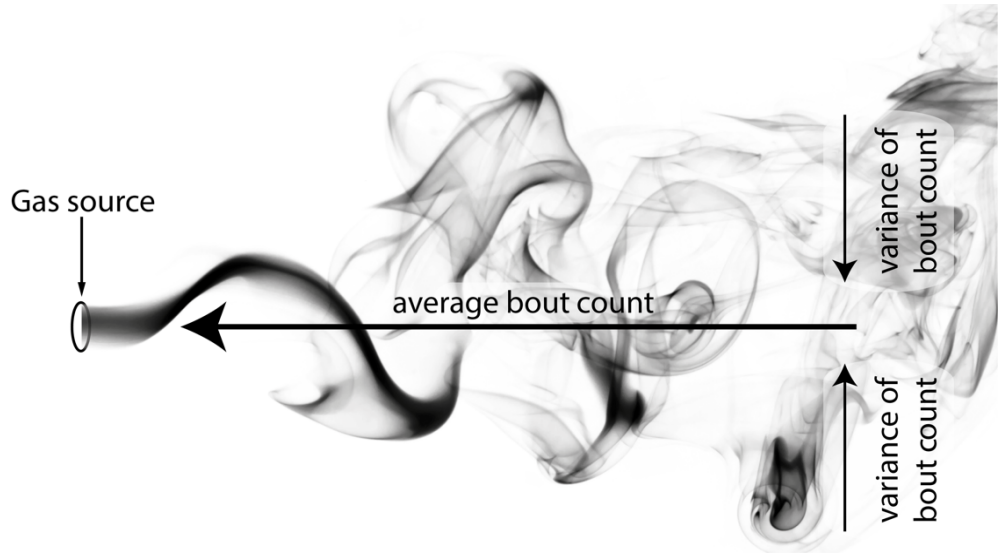

**Fig. 14:** Indicators for the location of a gas detector relative to the source. Downwind distance is encoded by average bout count, while a lateral displacement is encoded by the variance of bout counts.

These features give rise to a navigational strategy that could enable a mobile agent, e.g. a robot, to navigate towards a gas source based on the temporal features of a gas plume. The agent would first measure, at its current location, the number of gas bouts and their variance. If the variance of bouts per time is low, this indicates that the agent is located close to the centreline of the plume. Hence, it would move straight upwind for the next measurement. High variance of bout counts would indicate that the agent is located slightly off the centre of the plume. A good strategy would be to perform cross-wind casting moves in order to find the centre of the gas plume, from which it could start moving upwind.

In a practical scenario, it is preferable to assess the number of bout counts and their variance from continuous measurements, instead of a certain number of fixed-length measurement blocks. Ideally, the uncertainty of estimate of bout rate and variance would decrease with each bout that is encountered. This could be achieved by basing the statistics on inter-bout intervals (IBI) rather than bout counts. The average bout count $\langle n_{bout} \rangle$ can be replaced by $\frac{1}{\langle IBI \rangle}$, with $\langle IBI \rangle$ the average length of all inter-bout intervals that have been encountered so far. Similarly, the standard deviation of bout counts $std(n_{bout})$ could be replaced by $std$(IBI). Replacing bout count statistics with inter-bout interval statistics yields qualitatively very similar results (see supplemental Fig. S6). The advantage of inter-bout interval statistics is that they do not require to partition the measurement period into discrete chunks. Rather, they can be updated continuously, allowing for appropriate action to be taken as soon as enough bouts for a reliable estimate of mean and variance have been encountered.

Our results motivate future work on putative gas-based strategies for robot navigation. Clearly, the structure of turbulent gas plumes (and hence also the bout statistics) depend on numerous factors in the environment, as e.g. wind speed, shape of the terrain, the presence of buildings or other obstacles to airflow, among others. To further explore the effect of such factors, we think that a potential navigation strategy should initially be tested in a simulated environment, that combines the simulation of turbulent plume dispersal with gas sensing on robots via metal oxide sensors. Such an environment has recently been proposed by Khaliq and coworkers [22]. A simulated test setup will also facilitate initial comparisons to other gas-based navigation strategies such as Infotaxis and/or reactive strategies [23,24].

# 5 Conclusions

We show that it is possible to predict the distance of a gas source in a turbulent environment from the temporal structure of a MOX-sensor signal. The distance estimation was significantly enhanced by using a simple filtering approach that is suitable for implementation on microcontrollers. This finding demonstrates that the spatio-temporal structure of turbulent gas plumes is accessible with MOX-sensors, and that we can extract useful information from such measurements. Our findings may enable enhanced estimation of gas source distance, and potentially also source localisation by mobile agents using low-cost, low-power platforms.

# 6 Vitae

**Michael Schmuker** holds a degree in Biology (University of Freiburg, Germany, 2003), and a PhD in Chemistry (Goethe-University Frankfurt, Germany). His postdoctoral work focused on computational neuroscience of Olfaction (Freie Universität Berlin, Germany, 2007-2014). From October to November 2013 he was a Visiting Scholar with Ramon Huerta at the BioCircuits Institute, UC San Diego, who introduced him to the world of electronic gas sensors. In 2014, he joined the School of Engineering and Informatics, University of Sussex, Brighton, UK, as a Marie Curie Research Fellow. His research revolves around bio-inspired signal processing, neuromorphic computing and exploring the chemical domain of odour space.

**Viktor Bahr** was a Bachelor student in Bioinformatics working in Dr. Schmuker's group at Freie Universität Berlin, where he received his degree in 2015. He is now pursuing his Master's degree in Informatics at Technical University Berlin, Germany.

**Ramón Huerta** received his Ph.D. from Universidad Autónoma de Madrid in 1994. He is a Research Scientist at the BioCircuits Institute, UC San Diego. Prior his current appointment, he was Associate Professor at the Universidad Autónoma de Madrid (Spain). His areas of expertise include dynamic systems, artificial intelligence, and neuroscience. His work deals with the development algorithms for the discrimination and quantification of complex multidimensional time series, model building to understand the information processing in the brain, and chemical sensing and machine olfaction applications based on bio-inspired technology. Dr. Huerta's research work gathers in a publication record of over 100 articles in peer-reviewed journals at the intersection of computer science, physics, and biology.

# 7 Acknowledgements

Author contributions: MS conceived the study. VB and MS wrote analysis code and analysed data. RH provided data and support. MS wrote the manuscript. We thank Dr Thomas Nowotny for valuable comments on the manuscript. Gas plume image in Fig. 14 © William Warby (CC-BY 2.0). This study was supported by EU FP7 Marie Curie Intra-European Fellowship (project BIOMACHINELEARNING, grant no. 331892).

# Exploiting plume structure to decode gas source distance using metal-oxide gas sensors

– Supplemental information –

Michael Schmuker[1,2,3] *, Viktor Bahr[2], Ramón Huerta[3]

1: University of Sussex, School of Engineering and Informatics, Falmer, Brighton BN1 9QJ, UK.

2: Freie Universität Berlin, Dept. of Biology, Chemistry, Pharmacy, Königin-Luise-Str. 1-3, 14195 Berlin, Germany

3: University of California San Diego, BioCircuits Institute, La Jolla, CA 92093-0328, USA.

* corresponding author

## S1 Artefacts on Sensor 1

Sensor 1 has proven to be unusable on Board 5 due to strong artefacts in the recordings that could not be mitigated by post-processing. The signal contains "jumps" that occur inadvertently (Fig. S1A). These jumps occur on virtually all recordings of sensor 1. These jumps have a complex temporal structure (Fig. S1B). We were unable to pinpoint the cause for these jumps. Due to their complex structure it's difficult to remove them by post-processing without introducing other artefacts that might affect the analysis. Hence, we decided to exclude all recordings from sensor 1 from the analysis in this study.

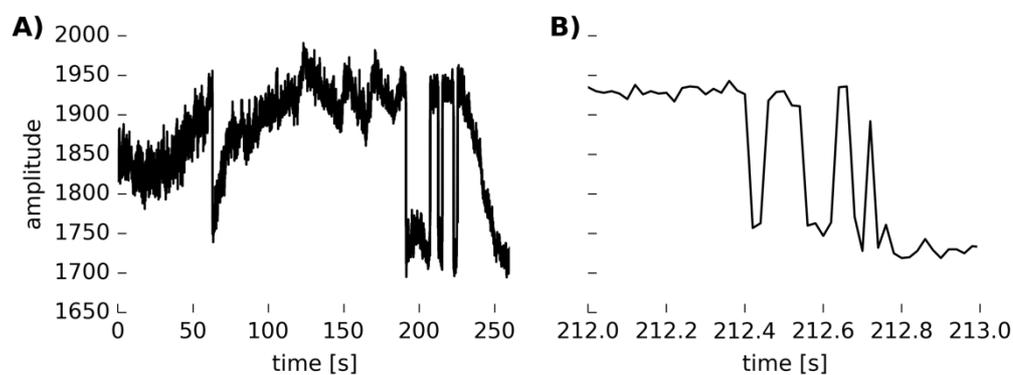

**Fig. S1:** Artefacts in the signal from sensor 1. The signal jumps between high and low values. A) A representative example of an entire measurement. B) One of the "jump events" in detail.

## S2 Influence of wind speed

The data set contains measurements at three different wind speeds. We analysed the influence of wind speed in the wind tunnel on the bout counts and the reliability of distance prediction. Fig. S2A shows the signal traces and detected bouts for the slowest wind speed (i.e. for, the lowest number of fan revolutions per minute, rpm), for two selected positions. As wind speed increases, the number of detected bouts goes down, as well as their amplitudes, as visible in single traces for the medium (Fig. S2B) and highest wind speed (Fig. S2C). While the bouts get weaker and fewer in number, the overall relationship between relative bout count and distance to source holds also for all three wind speeds (Fig. 2D, E, F). The precision of the distance prediction is hardly affected in cross-validation in this example. However, this observation indicates that a model that aims to predict source distance in an uncontrolled environment (as opposed to sampling in a wind tunnel with controlled wind speed) should also incorporate wind speed measurements into the predictive regression.

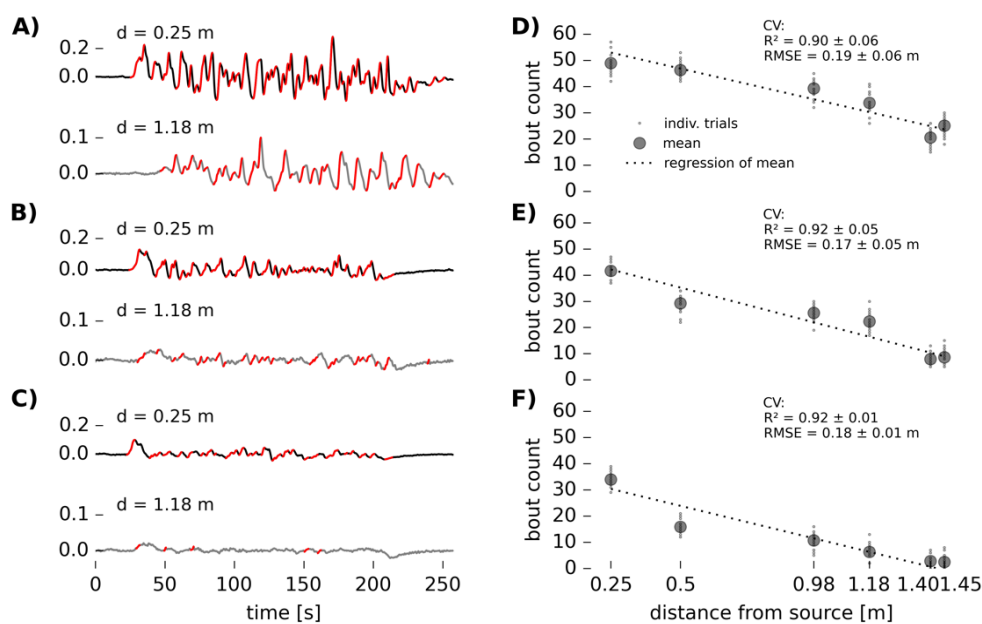

**Fig. S2:** Dependence on wind speed in the wind tunnel. A) Filtered trace and detected bouts for one trial at the slowest fan setting (1500 rpm), at two distances. Acetaldehyde 500 ppm, TGS 2610, trial number 10 (same as in Fig. 5A). B) Same gas, fan speed 3900 rpm, C) fan speed 5500 rpm. D) Regression for 20 trials with the fan speed 1500 rpm, E) fan speed 3900 rpm, F) 5500 rpm. Text insets in D,E,F show performance of cross-validated linear regression.

## S3 Influence of heater voltage

The original data was recorded using 6 heater voltages, from 4 to 6 V in steps of 0.5V. A higher heater voltage results in higher heater current and therefore higher sensor temperature. We analysed the influence of sensor voltage on the bout count feature and distance prediction. The effect of heater voltage is shown in Fig. S3. The signals acquired with heater voltages of 5V and 6V are qualitatively very similar, and the regression of bout counts vs distance yields similar accuracy in cross-validated prediction. At a heater voltage of 4V, the signal gets considerably weaker, resolving fewer bouts. Nevertheless, the RMSE of the cross-validated distance prediction is comparable to the performance obtained with the higher heater voltage.

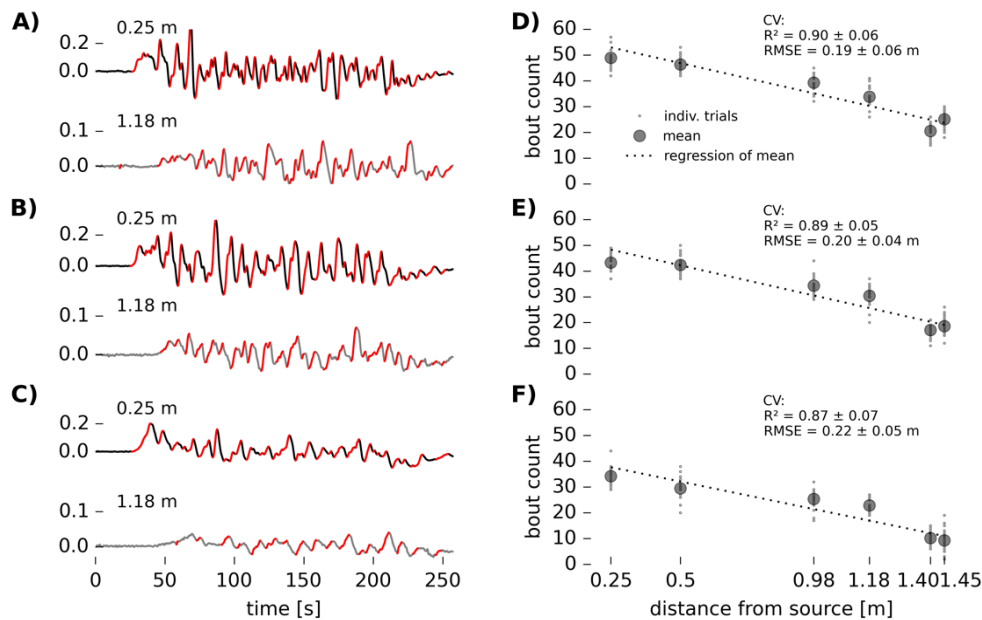

**Fig. S3:** Influence of heater voltage. A,B,C) EWMA-filtered signals and detected bouts for heater voltage 6V (A), 5V (B) and 4V (C). Acetaldehyde 500 ppm, trial number 1 (*cf.* Fig. 5 and Fig. S2), sensor TGS 2610. D,E,F) Bout counts vs. source distance for heater voltage 6V (D), 5V (E) and 4 V (F). Text insets show performance of cross-validated linear regression.

## S4 Similarity of bout timings across sensors

In order to quantify the similarity of bouts across sensors we adopt an approach first described by Schreiber et al (2003) that aims at measuring the similarity of event series. This approach is based on convolving a time series of discrete events (in the original study, neuronal action potentials) with a Gaussian kernel, thus creating a continuous time series. The similarity of two time-series is then quantified by the Pearson correlation of these continuous series.

Here, we apply this measure of event series to the bout onsets as discrete events. Fig. S4 depicts the bout onset times for the signals in Fig. 6 (Acetaldehyde, source distance 1.18 m, trial 19). We convolved these time series with Gaussian kernels of width $\sigma = 2\,s$. We then computed the pairwise correlation coefficients between the generated continuous time series. This analysis was done for all 20 trials that were present in the data set for Acetaldehyde, measured in 1.18 m distance from the source. The average correlation between all time series was $c = 0.38 \pm 0.21$ (standard deviation).

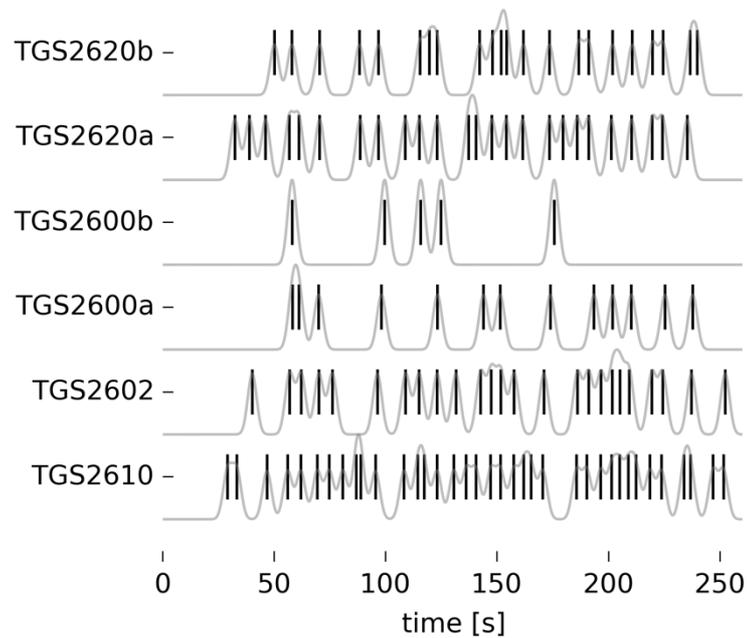

**Fig. S4**: Bout onset times as discrete events (black), and as smoothed representation (Gaussian kernel, $\sigma = 2\ s$, gray).

To check against a random background, we scrambled the trials, i.e., computing correlations between time series that were randomly chosen from different trials. Here we obtained $c = 0.20 \pm 0.14$. Fig. S5 depicts the histograms of pairwise correlations obtained in matched and randomised trials. A 2-sample Kolmogorov-Smirnov test confirmed that the correlations observed in pairs of matched trials is significantly different from randomised trials ($p = 2.9 \cdot 10^{-24}$).

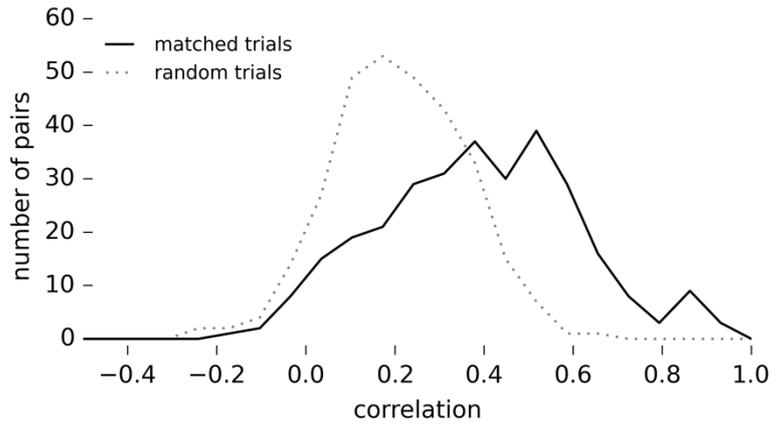

**Fig S5**: Statistics of pairwise correlations between smoothed time-series for matched trials (black) and trials that were randomly picked out of the 20 repetitions (gray).

## S5 Assessment of inter-bout interval statistics

In a scenario where an agent interacts with the world in real-time, it is impractical to use trial-based bout count statistics. Rather, it would be preferable to use accumulative statistics that can be obtained continuously. To this end, trial-based bout count statistics can be replaced by inter-bout interval statistics. The average bout count can be replaced by the average length of the inter-bout interval (IBI). The average IBI duration becomes smaller as source distance increases (Fig. S6A), although the measure is very sensitive to noise in when the total bout count is low, e.g. close to the walls of the wind tunnel (boards 1, 2, 3 and 7, 8, 9). Likewise, the standard deviation of bout counts over trials can be replaced by the standard deviation of inter-bout intervals in a trial, which is minimal near the centreline of the plume (Fig. S6B). Normalising the standard deviation by the average IBI yields gives even clearer minima close to the centreline of the plume for all distances, potentially enabling more robust navigation (Fig. S6C). Taken together, the IBI statistics exhibit a similar relationship to down- and crosswind distance to the source, but with the advantage that they can be obtained in an accumulative, continuous fashion, and could replace trial-based approaches in interactive scenarios.

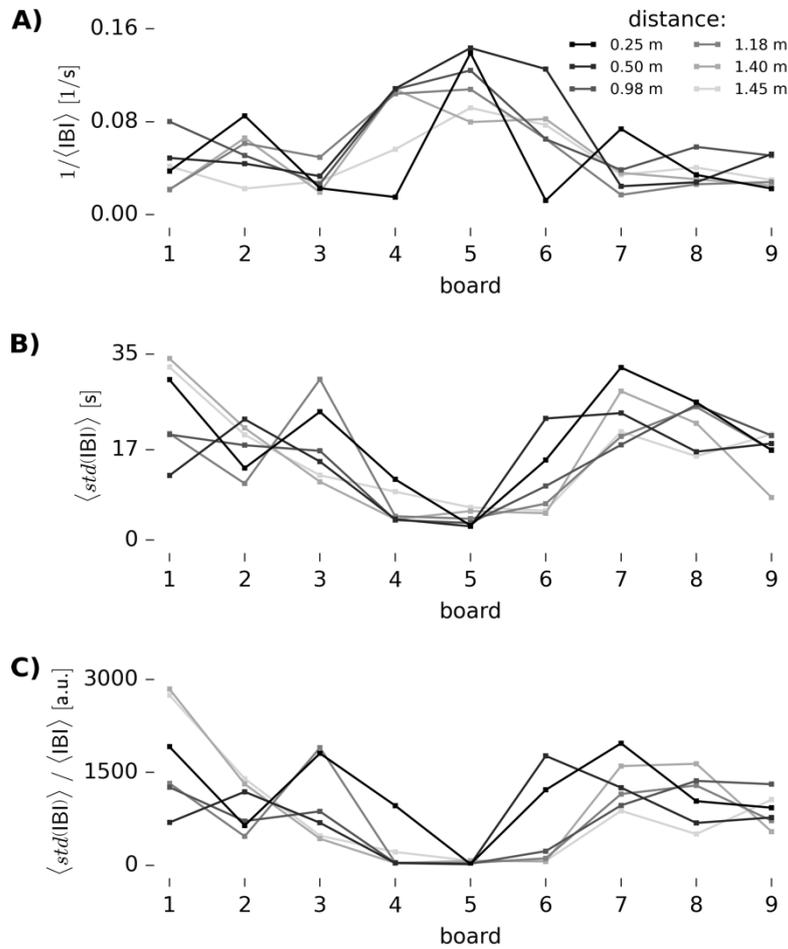

**Fig. S6:** Inter-bout interval (IBI) statistics in down- and crosswind direction from the source. Board number refers to cross-wind position of sensors, with board no. 5 at the centre. Distance from source is indicated by grey-scale. A) Reciprocal average duration of inter-bout interval. Measurements with less than 3 bouts were ignored. B) Standard deviation of inter-bout intervals (average across trials). C) Average standard deviation normalised by average inter-bout interval.

## Supplemental References